\documentclass[12pt]{article}
\usepackage[square,comma,sort&compress]{natbib}
\usepackage{helvet,times,mathptm,amsmath,amssymb}
\usepackage{epsfig,bbm,amsfonts}
\usepackage{graphicx,color}
\newcommand\beq{\begin{equation}}
\newcommand\eeq{\end{equation}}
\newcommand\bea{\begin{eqnarray}}
\newcommand\eea{\end{eqnarray}}

\newcommand\bra{\langle}
\newcommand\ket{\rangle}

\begin{document}

\begin{center}
{\noindent
{\bf Real Time Dynamics of Hole Propagation in Strongly \\
Correlated Conjugated Molecular Chains: A time-dependent DMRG Study}} 
\end{center}

\begin{center}
{\noindent
{\small \bf Tirthankar Dutta \footnote[1]{tirthankar@sscu.iisc.ernet.in},
and \bf S. Ramasesha \footnote[2]{ramasesh@sscu.iisc.ernet.in} }}
\end{center}

\begin{center}
{\noindent
{\small {\it
Solid State $\&$ Structural Chemistry Unit,\\
Indian Institute of Science,\\
Bangalore-560012, INDIA.}}}
\end{center}

\vspace{0.2cm}

\centerline{\bf Abstract}
~\\
{\noindent {\small
In this paper, we address the role of electron-electron interactions
on the velocities of spin and charge transport in one-dimensional
systems typified by conjugated polymers. We employ the Hubbard model to
model electron-electron interactions. The recently developed technique
of time dependent Density Matrix Renormalization Group (tdDMRG) is used
to follow the spin and charge evolution in an initial wavepacket
described by a hole doped in the ground state of the neutral system. We
find that the charge and spin velocities are different in the presence
of correlations and are in accordance with results from earlier
studies; the charge and spin move together in noninteracting picture
while interaction slows down only the spin velocity. We also note that
dimerization of the chain only weakly affects these velocities.}}

\section{Introduction}
{\noindent
Low-dimensional many-body systems have always been the focus of theoretical and experimental
interest. The physics of these systems is quite different from those of three (3D) systems. For 
example, these materials show the phenomena of spin-charge separation, wherein the spin and charge
degrees of freedom of the electron get decoupled and evolve independently of each other with 
different velocities. These materials find wide scale applications in the field of molecular 
electronics (spintronics). Amongst low-dimensional materials, the $\pi$-conjugated polymers have 
attracted a lot of interest, being potential candidates for various molecular electronics and 
spintronics applications; examples include the organic light emitting diodes (OLEDS), 
organic semiconductors, organic thin film transistors, etc. \cite{dodabalapur,katz,bradeley,nitzan}. 
However, spin and charge transport in these systems is still not well established because of 
the strong electron-electron correlations that exist in these systems. Transport in these
materials is strictly a non-equilibrium phenomena to understand which, one needs to investigate the 
time evolution of strongly interacting quantum many body systems. Recently, there has been a 
considerable progress in investigation of non-equilibrium time evolution of many body systems. 
Analytical approaches like the perturbative Keldysh formalism \cite{keldysh}, is restricted to a few 
integrable models only, but in the case of low-dimensional systems, efficient numerically accurate
techniques have been developed and successfully applied to a variety of models. One
such efficient method which has gained tremendous impetus in recent years, is the 
time-dependent Density Matrix Renormalization Group technique (tdDMRG) 
\cite{dmrg1,dmrg2,dmrg3,dmrg4,dmrg5,dmrg6}.  
In this paper, we use tdDMRG to address the effects of (1) electron-electron correlations and, (2)
dimerization, on the charge (spin) transport in quasi 1-dimensional strongly 
correlated polyene chains. Moreover, we also look into the dynamics of spin-charge separation in these 
systems. To address the above questions, we focus our attention on the real time quantum dynamics of 
an hole with up spin injected at site-1 of polyene chains.} 

\section{Model Hamiltonians and Parameters}
{\noindent
We have modeled the $\pi$-conjugated chains using three model Hamiltonians: (a) the 
Tight-Binding (TB) Hamiltonian, known as H\"uckel model to chemists \cite{huckel1,huckel2}, 
(b) the single-band Hubbard model \cite{hubbard1,hubbard2,hubbard3},
and (c) Pariser-Parr-Pople (PPP) model \cite{ppp1,ppp2}. 
Amongst these, 
(2) and (3) are interacting Hamiltonians that include explicit electron correlations and
(3) is a realistic model Hamiltonian used for 
describing $\pi$-conjugated polymers. In second quantized formalism, these three model Hamiltonians can 
be expressed as given below \cite{huckel3}:}
\beq
H_{\text{H\"uckel}} = t_{0} \sum^{N-1}_{i=1, \sigma=\uparrow,\downarrow}(a^{\dagger}_{i+1 \sigma}a_{i \sigma}
+ h.c) 
\eeq  
\beq
H_{Hubbard} = H_{\text{H\"uckel}} + U \sum_{i=1}^{N} n_{i \uparrow}n_{i \downarrow}
\eeq
\beq
H_{PPP} = H_{Hubbard} + \sum_{i>j}V_{ij}(n_{i}-z_{i})(n_{j}-z_{j})
\eeq
\newline
{\noindent
Here, $a^{\dagger}_{i \sigma}$ ($a_{i \sigma}$) creates (annihilates) an electron at site-i of the
polyene chain, $t_{0}$ is the  nearest-neighbour (nn) hopping integral for an undimerized chain, and 
h.c refers to the hermitian conjugate. In the case of a dimerized polyene chain, the nn hopping 
integral is given by, $t_{i,i+1} = t_{0} (1-(-\delta)^{i})$, where $\delta$ is the dimerization 
parameter. For the present study, we have taken $\delta$ = 0.07, so that the nn hoping term for long 
and short bonds are respectively given by, $t_{long-bond}$ = 1.07 and $t_{short-bond}$ = 0.93, 
$t_{0}$ = 1.0 for the Hubbard model. $n_{i \uparrow}$ ($n_{i \downarrow}$) are the number density of 
upspin (downspin) electrons at site-i of the polyene chain. The Hubbard model is characterized by $U$, 
the Hubbard parameter, which represents \underline{on-site} Coulomb repulsion between two electrons of 
opposite spins occupying the same site of the polyene chain. For homogenous systems, this parameter is 
same for all sites. $U$ is measured in terms of $t_{0}$, and the parameter, ($U/t_{0}$) characterizes 
the $\pi$-electronic motion in single band systems. In our study, we have taken $U/t_{0}$ = 0.0 
(the H\"uckel model), 2.0, 4.0, 6.0 and 10.0. In the PPP model, the $V_{ij}$ is the inter-site Coulomb 
repulsion between two different sites, i and j, of the polyene chain. In keeping with the spirit of 
phenomenology associated with the PPP Hamiltonian, the inter-site electron repulsion integrals, 
$V_{ij}$ are interpolated smoothly between $U$ for zero separation and $\frac{e^{2}}{r_{12}}$ for the 
inter-site separation tending to infinity; thus, the explicit evaluation of the repulsion integrals is 
avoided. There are two widely used interpolation schemes used to evaluate $V_{ij}$, the Ohno scheme 
\cite{ohno}, and the Mataga-Nishimoto scheme \cite{mataga}. In the Ohno interpolation scheme which we 
use, the inter-site electron repulsion integrals, $V_{ij}$ are given by,}
\beq
V_{ij} = 14.397 \biggl[ \biggl(\frac{28.794}{U_i+U_j} \biggr)^2 + r^{2}_{ij} \biggr]^{-1/2}
\eeq  
\newline
{\noindent
The Mataga-Nishimoto formula is given by,}     
\beq
V_{ij} = \biggl[ \frac{2.0}{U_i+U_j} + \frac{r_{ij}}{14.397} \biggr]^{-1}
\eeq
\newline
{\noindent
The Ohno interpolation formula decays more rapidly than that of Mataga-Nishimoto scheme. In both the 
above interpolation schemes, it is assumed that $r_{ij}$ is in $\AA$, while $t_{0}$, $U$, and $V_{ij}$ 
are in measured in ev. $z_{i}$ is the chemical potential of site-i; the function of $z_{i}$ is to keep 
the i$^{th}$-carbon atom neutral when singly occupied.} 

\section{Time-dependent DMRG - Xiang's Algorithm}
{\noindent 
For carrying out quantum dynamics of the an up spin hole injected at site-1 of the polyene chain, we
first create the necessary initial state, by annihilating an up spin electron from site-1 from the ground state of
the $\pi$-conjugated chain. Mathematically, this amounts to the following:}
\beq
\mid \psi(0) \ket = a_{1 \uparrow} \mid \psi_{GS} \ket
\eeq
\newline
{\noindent
Here, $\mid \psi(0) \ket$ is the desired initial state and $\mid \psi_{GS} \ket$ is the ground state
of the neutral polyene chain. Using $\psi(0)$, we numerically solve the time-dependent Schr\"odinger 
equation (TDSE) which is given by,}
\beq
i\hbar \frac{\partial \mid \psi(t) \ket}{\partial t} = H \mid \psi(t) \ket
\eeq
\newline
{\noindent
where H is any of the three time-independent model Hamiltonians discussed in sec. II. The above equation
has the formal solution,} 
\beq
\mid \psi(t) \ket = e^{-iHt} \mid \psi(0) \ket 
\eeq
\newline
{\noindent
Numerically, given a small time step $\Delta t$, H and $\mid \psi(0) \ket$, the TDSE can be solved by expanding the 
exponential function in equation (8), to various orders of (H $\Delta t$). The simplest of this is
the Euler (EU) scheme as
given below:}
\beq
\mid \psi(t+\Delta t) \ket = (1 - iH\Delta t) \mid \psi(t) \ket + O((H \Delta t)^{2})
\eeq
\newline
{\noindent  
This equation is then repeatedly used to obtain to propagate the initial wave-packet. This scheme is an 
explicit one without requiring any matrix inversion. However, it suffers from two serious drawbacks:
(1) it is non-unitarity, and (2) there is an instability due to the lack of time inversion symmetry (t $\longrightarrow$
-t). To avoid these problems, the TDSE is solved using the implicit Crank-Nicholson (CN) scheme \cite{cn} 
in which the exponential function is approximated by the Caley transform}
\beq
\mid \psi(t+\Delta t) \ket = \frac{1 - iH\Delta t/2}{1 + iH\Delta t/2} \mid \psi(t) \ket + 
O((H\Delta t)^3)
\eeq
\newline
{\noindent
The CN scheme is unitary, unconditionally stable, and accurate up to $(H\Delta t)^3$. However, this
scheme also has a serious limitation, namely: Each time evolution step requires a matrix 
inversion, which for large systems and with the increase of dimensionality, requires huge memory and 
CPU time, 
making this method prohibitive. Hence, there has been a surge towards the development of 
explicit, stable integration schemes. The first of these is a symmetrized  version of the EU scheme,
called the second order differencing scheme (MSD2) \cite{askar}. This scheme is symmetric in time
as seen below, is conditionally stable, and accurate upto $(H \Delta t)^2$.}
\beq
\mid \psi(t+\Delta t) \ket = -2iH\Delta t \mid \psi(t) \ket + \mid \psi(t-\Delta t) \ket + 
O((H\Delta t)^3)
\eeq 
\newline
{\noindent
The MSD2 scheme can be extended to higher order accuracy forms, which are collectively called the 
{\it multistep differencing} (MSD) schemes \cite{iitaka}, for example, the fourth and sixth order MSD
(MSD4 and MSD6) which can given as below (equations (12) and,(13)):}
\newline
\beq
\begin{split}
\mid \psi(t+2\Delta t) \ket &= \mid \psi(t-2\Delta t) \ket + 4iH\Delta t \biggl[ \frac{1}{3}\mid 
\psi(t) \ket -\frac{2}{3} \biggr( \mid \psi(t+\Delta t) \ket + \mid \psi(t-\Delta t) \ket \biggr) 
\biggr ] \\
                            &+ O((H\Delta t)^5)
\end{split}
\eeq
\beq
\begin{split}
\mid \psi(t+3\Delta t) \ket &= \mid \psi(t-3\Delta t) \ket - 6iH\Delta t \biggl[ \frac{13}{10}\mid 
\psi(t) \ket -\frac{7}{10} \biggl( \mid \psi(t+\Delta t) \ket + \mid \psi(t-\Delta t) \ket \biggr) \\
                            &+ \frac{11}{20} \biggl( \mid \psi(t+ 2\Delta t) \ket + \mid 
\psi(t-2\Delta t) \ket \biggr) \biggr ] + O((H\Delta t)^7) 
\end{split}
\eeq 
\newline
{\noindent
These higher order schemes are explicit and conditionally stable, for example, MSD4 is stable if and
only $\Delta t <$ 0.4, while for MSD6, stability exists for $\Delta t <$ 0.1. Predictor-Corrector 
(PC) techniques are another class of ordinary differential equation solvers. For our present studies, 
we have developed a PC scheme of our own, which we call the {\it MSD4-AM4 method}. In this, we use the 
explicit MSD4 (equation (12)) scheme as the predictor, and fourth order implicit Adams-Moultan method 
as the corrector (equation (14)). We found this scheme to be very robust, and as efficient as the CN 
method; moreover, this PC technique is much faster and less memory consuming 
compared to the CN scheme \cite{tirthankar}.}
\newline
\beq
\begin{split}
\mid \psi(t+2\Delta t) \ket &= \mid \psi(t+\Delta t) \ket -\frac{iH\Delta t}{24} \biggl( 
9\mid \psi(t+2\Delta t) \ket + 19\mid \psi(t+\Delta t) \ket - 5\mid \psi(t) \ket \\ 
                            &+ \mid \psi(t-\Delta t) \ket \biggr) + O((H\Delta t)^5)
\end{split}
\eeq
\newline
{\noindent
So far we have discussed about model Hamiltonians, preparing the initial state, and time evolution of 
this state by solving the TDSE numerically. For obtaining the initial and ground states of the polyene 
chains, we use tdDMRG as given by Xiang and coworkers \cite{xiang}. However, before discussing this 
technique, we'll briefly discuss the conventional {\it infinite system} DMRG method \cite{dmrg1,dmrg2} 
as proposed by White and others. The basic idea of the DMRG method is to divide a given finite system
into two parts, namely, {\it system} and {\it surrounding}, followed by retaining only the {\it m} most
highly weighted eigenstates of the reduced density matrix of these partial "systems" \cite{dmrg7}. 
Using these reduced density matrices, one or more pure states of this total system is obtained. 
In case of the {\it infinite system} DMRG algorithm, the system size is increased in units of 
"two sites" (see fig. 1) \cite{dmrg1,dmrg2,dmrg3}.}
\vspace{0.2cm}
\newline

\begin{figure}[!htbp]
\begin{center}
{\includegraphics[scale=0.51]{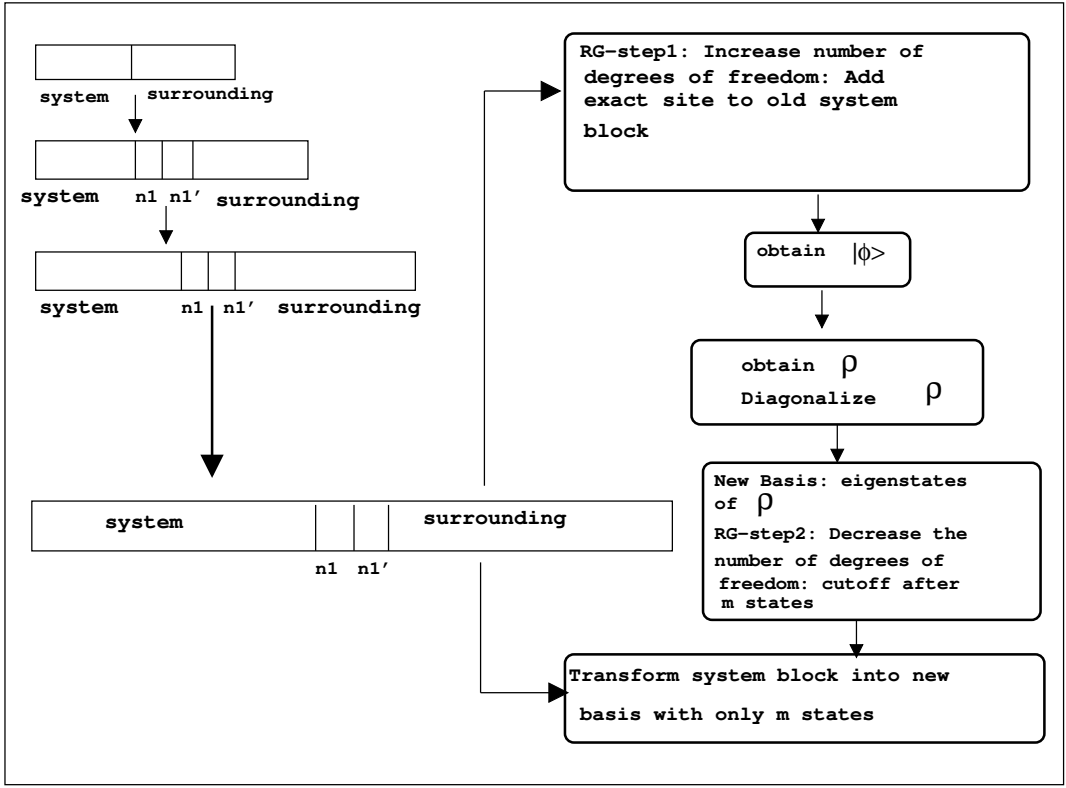}}
\caption{\small
Schematic sketch of infinite system DMRG algorithm along with the flowchart 
showing basics of the DMRG truncation scheme. For more details of the method, see \cite{dmrg2,dmrg3}. 
n1 and n1' are the two new sites that are attached at each infinite system DMRG step.}
\label{figure-1} 
\end{center}
\end{figure}

\newpage
{\noindent
So far, all tdDMRG schemes can be categorized into three classes: (1) Static time-dependent DMRG, 
(2) Dynamic time-dependent DMRG, and (3) Adaptive time-dependent DMRG method \cite{dmrg4}. The static 
tdDMRG method was first introduced by Cazalilla and Marston \cite{marston}, who exploited this 
technique for investigating time-dependent quantum many-body effects. They studied a time-dependent 
Hamiltonian, H(t) = H(0) + V(t), where V(t) represents the time-dependent part of the Hamiltonian.
Initially, infinite system DMRG method was used for constructing a lattice of desired size keeping a 
substantially large number of reduced density matrix eigenstates ({\it m}). 
Time evolution of this final 
lattice system is then carried out using the time-dependent Hamiltonian, $H_{eff}$(t), which is given 
by, $H_{eff}$(t) = $H_{eff}$(0) + $V_{eff}$(t), where $H_{eff}$(0) is the final superblock Hamiltonian 
approximating H(0), and $V_{eff}$(t) is an approximation of V(t), and is built using the 
representations of operators in the final block bases. The basic idea of this method is to fix the
reduced Hilbert space at its optimal value at time t = 0, and then, projecting 
all wavefunctions and operators
on to it. In other words, the effective Hamiltonian which has been obtained by targeting the
ground state of the t = 0 Hamiltonian is capable of representing adequately the time-dependent states
that will be reached at later times. The major disadvantage of this scheme is that it fails
completely for long time evolution as there is a significant loss of information due to the 'final 
superblock truncation'. Moreover, the number of DMRG
states, {\it m}, grows with the simulation time as they need to incorporate a constantly increasing number of 
nonequilibrium states. To overcome this, in 2003, Luo, Xiang and Wang \cite{xiang} came up with a 
targeting method, which is called the Dynamic tdDMRG or the LXW method, and will be utilized by us, for the present 
study. We will however, not discuss the Adaptive tdDMRG scheme. Interested readers can
refer the relevant articles \cite{dmrg4,dmrg5,dmrg6,adap1,adap2,adap3}. 
The algorithm, as implemented by us, is given in details below:
\newline
{\noindent
(1) The Hamiltonian of a small, exactly diagonalizable superblock(SB) of {\bf L} (= 4) sites, $H^{SB}_{L=4}$,  is first constructed.}
\newline
{\noindent
(2) The ground state, $\psi_{gs}^{4}$, of this 4-site SB is obtained by exact diagonalization of 
$H^{SB}_{L=4}$. Using $\psi_{gs}^{4}$, a desired initial state $\psi_{0}^{4}$ of interest is prepared.
Exact time evolution of this initial state is then carried out from t = 0, to t = $N_{steps}$, by
solving the TDSE numerically, using a convenient integration scheme. (In our case, we use our own
MSD4-AM4 scheme). At the end of the time evolution, a set of time-dependent wavefunctions are obtained,
$\{ \psi(t_i): ~t_i ~\forall \in ~(0, N_{steps}) \}$. }
\newline
{\noindent
(3) Using this set of time-dependent wavefunctions, the reduced density matrices for the left-
($\rho_{l}$) and right-half ($\rho_{r}$) blocks for the next SB is build using LXW prescription
\cite{xiang}. Mathematically,}
\beq
\rho_{l} = Tr_{r} \biggl( \sum_{i=0}^{N_{steps}} \omega_{i} \mid \psi(t_i) \ket \bra \psi(t_i) \mid
\biggr); ~~~ \sum_{i=0}^{N_{steps}} \omega_{i} = 1
\eeq
\beq 
\rho_{r} = Tr_{l} \biggl( \sum_{i=0}^{N_{steps}} \omega_{i} \mid \psi(t_i) \ket \bra \psi(t_i) \mid 
\biggr); ~~~ \sum_{i=0}^{N_{steps}} \omega_{i} = 1
\eeq
\newline
{\noindent
Here, $\mid \psi(t_i) \ket$ is called the ith-target state and $\omega_i$ is its corresponding weight
in the half-block reduced density matrix. In the original LXW method, in building of $\rho_{l}$ and
$\rho_{r}$, only $\psi(0)$ and $\psi(t_i)$ $\forall$ i $\in$ $(1, N_{steps})$ are included. However, in
our case, we have two systems at hand: neutral polyene, and "cationic" polyene, having +1 charge on it.
We found that in the case of homogeneous systems, devoid of heteroatoms, it is un-important whether
we keep the "ionic" ground state for building of the density matrices. However, for systems with
heteroatoms, including the "ionic" ground state is {\it essential} for building the density matrices.
Furthermore, we have also found that by comparing tDMRG results with exact results, for small chains, $\omega_{ion}$ $<$ $\omega_{0}$ is required.
Hence, the above pair of equations get modified as,}
\beq
\rho_{l} = Tr_{r} \biggl( \omega_{gs,ion} \mid \psi_{gs,ion} \ket \bra \psi_{gs,ion} \mid + 
\sum_{i=0}^{N_{steps}} \omega_{i} \mid \psi(t_i) \ket \bra \psi(t_i) \mid \biggr); 
~~~ \omega_{gs,ion} + \sum_{i=0}^{N_{steps}} \omega_{i} = 1
\eeq
\beq
\rho_{r} = Tr_{l} \biggl( \omega_{gs,ion} \mid \psi_{gs,ion} \ket \bra \psi_{gs,ion} \mid + 
\sum_{i=0}^{N_{steps}} \omega_{i} \mid \psi(t_i) \ket \bra \psi(t_i) \mid \biggr);
~~~ \omega_{gs,ion} + \sum_{i=0}^{N_{steps}} \omega_{i} = 1 
\eeq
\newline
{\noindent 
(4) These reduced density matrices are then diagonalized using a dense matrix diagonalization routine
to obtain {\it m} eigenvectors, with largest eigenvalues. These eigenvector constitute the Density
Matrix Eigen-Vector (DMEV) basis.}
\newline
{\noindent
(5) $H_{l+1}$ ($H_{r+1}$) and other operators ($A_{l+1}$) are then constructed 
in the new system block and
transformed to the reduced DMEV basis using the transformations, 
$\tilde{H}_{l+1}$ = $O_{L}^{\dagger}H_{l+1}O_{L}$,
~$\tilde{A}_{l+1}$ = $O_{L}^{\dagger}A_{l+1}O_{L}$. Here, $O_{L}$ is a {\it (4m $\times$ m)} matrix whose columns
contain {\it m} highest eigenvectors of $\rho_{l}$ ($\rho_{r}$), and $A_{l+1}$ is an operator in the
system block (left-, or right-half block).}
\newline
{\noindent
(6) A new SB of size ({\bf L+2}) is formed, using $\tilde{H}_{l+1}$, two newly added sites, and 
$\tilde{H}_{r+1}$.}
\newline
{\noindent
(7) The steps from (2) to (6) are repeated to iteratively increase the SB size by two sites at a time.}
\vspace{0.3cm}
\newline
{\noindent
Apart from the "ionic ground state" correction to the original LXW Algorithm, we have also 
introduced another modification, which we call the "n-slot" modification. This modification basically
means that instead of storing all the time-dependent non-equilibrium wavefunctions, after every 
"n-th" time step, the wavefunction is stored for building the density matrix. We have studied "n" = 
10, 100, 500 and 1000 cases. It is found that for getting correct results using LXW Dynamic tdDMRG 
technique, we need "n" $>$ 25. The basic idea behind this method is that, {\it the time-dependent 
wavefunctions for a SB of size {\bf L}, explores the Hilbert space as much as possible, and transfers 
the information through the reduced density matrix, towards building Hilbert space of a SB of size,
({\bf L+2})}. However, this technique suffers from a major problem, namely, it needs large CPU times. 
Parallelizing this algorithm would mitigate this drawback.
The dynamical variables that we study are charge (spin) densities at site-1, site-L of the polyene 
chain, along with charge (spin) velocities. These variables will be discussed in detail in the next
section.} 
\vspace{0.5cm}   
\newline

\section{Results and Discussion}
{\noindent
In the previous section, dynamical variables that are calculated in this paper, were mentioned.
Here, we discuss these quantities in detail, along with our results. The charge (spin) density at the 
i$^{th}$-site of a given polyene chain, at time $\tau$ is given by:}
\beq
\bra n_{i}(\tau) \ket = \bra \psi(\tau) \mid (n_{i \uparrow} + n_{i \downarrow}) \mid \psi(\tau) \ket   
\eeq 
\beq
\bra S_{i}^{z}(\tau) \ket = \bra \psi(\tau) \mid (n_{i \uparrow} - n_{i \downarrow}) \mid \psi(\tau) \ket   
\eeq 
\newline
{\noindent
We have calculated these two quantities at all sites of a given polyene chain. However, we focus our 
attention on $\bra n_{L}(\tau) \ket$ ($\bra S^{z}_{L}(\tau)$), "L" being the last site of the polyene 
chain. We have considered chains containing 10, 20, 30 and 40 C-atoms in the present study. Two 
different evolution times have been used namely, 33 fs (femtoseconds) and 10 fs. The dimension of 
the DMEV Basis is kept at an optimal value of 200 for the H\"uckel and Hubbard chains. Site-L of 
the polyene chain at time t = 0 has $\bra n_{L} \ket$ = 1.0, with equal probability for being
occupied by either an "up (down) spin", thereby making $\bra S^{z}_{L} \ket$ = 0.0. As time 
progresses, the injected hole propagates from the first to the last site, and this is represented 
by appearance of a minima in the time evolution profiles of both charge and spin densities. The time 
at which the 1st minima appears is therefore the time taken by the injected hole to reach the other 
of the chain. Hence, we focus our attention on this quantity throughout our studies.}
\vspace{0.5cm}
\newline
\subsection{Dynamics in H\"uckel Chains:}
{\noindent
Fig. 2 shows the time evolution of charge (spin) densities at the last site of polyene chains of 
different lengths, governed by the H\"uckel Hamiltonian. The left- and right-plot depicts charge 
and spin density dynamics respectively. Solid curves are used for undimerized (regular) chains, 
while dashed curves, for the dimerized chains, with $\delta$ = 0.07. In case of H\"uckel chains, 
from fig. 2 it is seen that with increasing chain length, the time taken by the injected hole to 
reach the end of the chain also increases. The velocity of the hole appears to be reasonably constant
for systems of different length. Furthermore, dimerization appears to slightly decrease the 
"hole-velocity" compared to the uniform chain. Careful examination of the time profiles of charge 
(spin) densities also reveal that they are identical, in features indicating  that there is 
no {\it spin-charge separation}.} 

\vspace{0.2cm}
\begin{figure}[!htbp]
\begin{center}
{\includegraphics[scale=0.6]{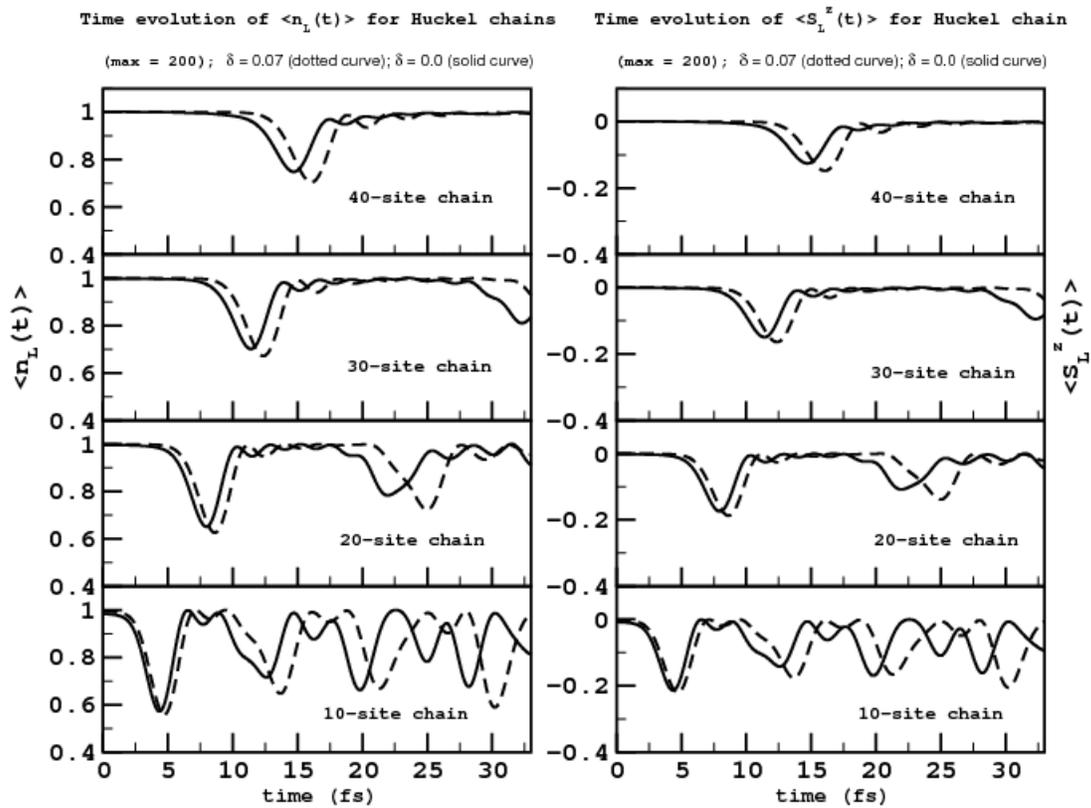}}
\caption{\small 
Time evolution of $\bra n_{L}(t) \ket$ ({\bf Left}) and $\bra S^{z}_{L}(t) \ket$ ({\bf Right}) for uniform 
(solid curve) and dimerized (dashed curve) H\'uckel chains of length 10, 20, 30 and 40 sites. 200 DMEV 
states per block is used.}
\label{figure-2} 
\end{center}
\end{figure}

\newpage
\subsection{Dynamics in Hubbard Chains:}
{\noindent
Figs. 3, 4, and 5 gives the time evolution profiles of charge and spin densities at the last site of 
Hubbard chains for different chain lengths, and for several representative values of U, namely, 
U/t = 2.0, 4.0, and 6.0, respectively. It is observed that the charge and spin dynamics are no longer 
identical as was seen in case of H\"uckel chains clearly indicating spin-charge separation. Furthermore,
as the magnitude of U increases, the extent of spin-charge decoupling also increases. In the 
literature, the decoupled spin and charge excitations are referred to as {\it spinons}, and 
{\it holons} \cite{kivelson,zou}. For a given U/t, "velocity" of the charge excitation (holon) as 
well as that of the spin excitation (spinon) seem to be weakly dependent on the chain length. It is 
however that at all chain lengths and nonzero U/t values, "velocity" of the holon is higher than that 
of the spinon. If one examines the plots carefully, another interesting observation can be made; in the 
correlated models, dimerization plays little or no role in influencing the "velocity" of the injected hole.
Figs. 6 and 7, show the variation of $\bra n_{L}(\tau) \ket$ and $\bra S_{L}^{z}(\tau) \ket$ 
for regular and dimerized chains of given length, for different values of U. Solid curves are 
for U = 2.0, dashed curves represents U = 4.0, while U = 6.0 is depicted by dotted curves. 
It is observed that for a fixed chain length, increasing U, does not perceptibly affect 
the velocity of the holon appreciably, but spinon's velocity is considerably altered. This
is simply because, in case of the 1-dimensional Hubbard model, analytical expressions for 
the holon ($v_{h}$) and spinon ($v_{s}$) velocities are given by \cite{karen1,coll},}
\newline
\beq
v_{h} = 2t \sin{(\pi n)}; ~~v_{s} = \frac{2 \pi t^{2}}{U} \biggl[1 - \frac{\sin{(2 \pi n)}}{2 \pi n} \biggr]
\eeq
\newline
{\noindent
where, $t$ and U are the nn hopping matrix element and the one-site Coulomb repulsion term,
respectively, and $n$ is the particle density (n $\le$ 1). Clearly, the velocity
of the holon does not depend on U, while that of the spinon decreases, as established also from 
our tdDMRG studies. Furthermore, as U $\rightarrow \infty$, spin velocity goes to zero. 
The holon moves by virtue of the hopping matrix element while the effective spin-spin exchange, 
which is of the order of $\frac{t^2}{U}$,  propagates the spinon. And, in the thermodynamic limit, 
that is, U $\rightarrow \infty$ limit, only the holon propagates, the spinon doesn't "move" at all, 
$v_{s}$ being zero. As the magnitude of U increases, the velocity of the spinon decreases.}

\vspace{0.2cm}
\begin{figure}[!htbp]
\begin{center}
{\includegraphics[scale=0.6]{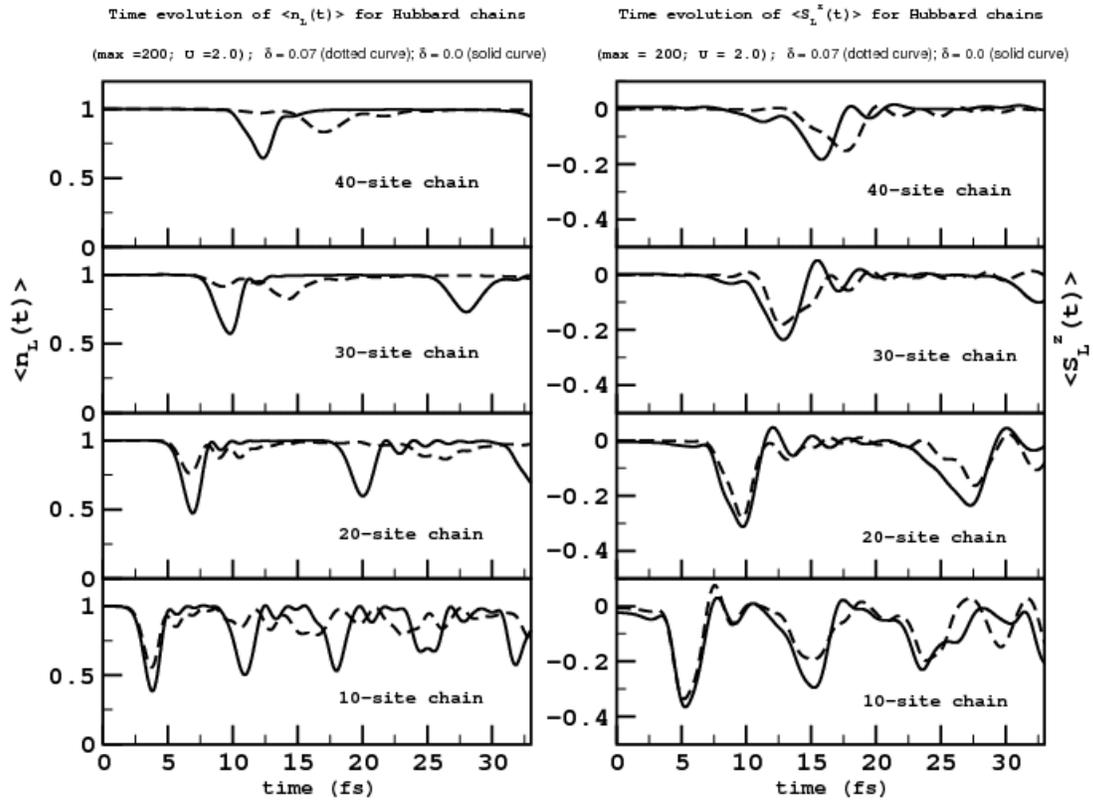}}
\caption{\small
Time evolution of $\bra n_{L}(t) \ket$ ({\bf Left}) and $\bra S^{z}_{L}(t) \ket$ ({\bf Right}) for uniform
(solid curve) and dimerized (dashed curve) Hubbard chains of length 10, 20, 30 and 40 sites, with
$\frac{U}{\mid t \mid}$ = 2.0. Dimension of the DMEV basis used is 200.}
\label{figure-3}
\end{center}
\end{figure}

\vspace{0.2cm}
\begin{figure}[!htbp]
\begin{center}
{\includegraphics[scale=0.6]{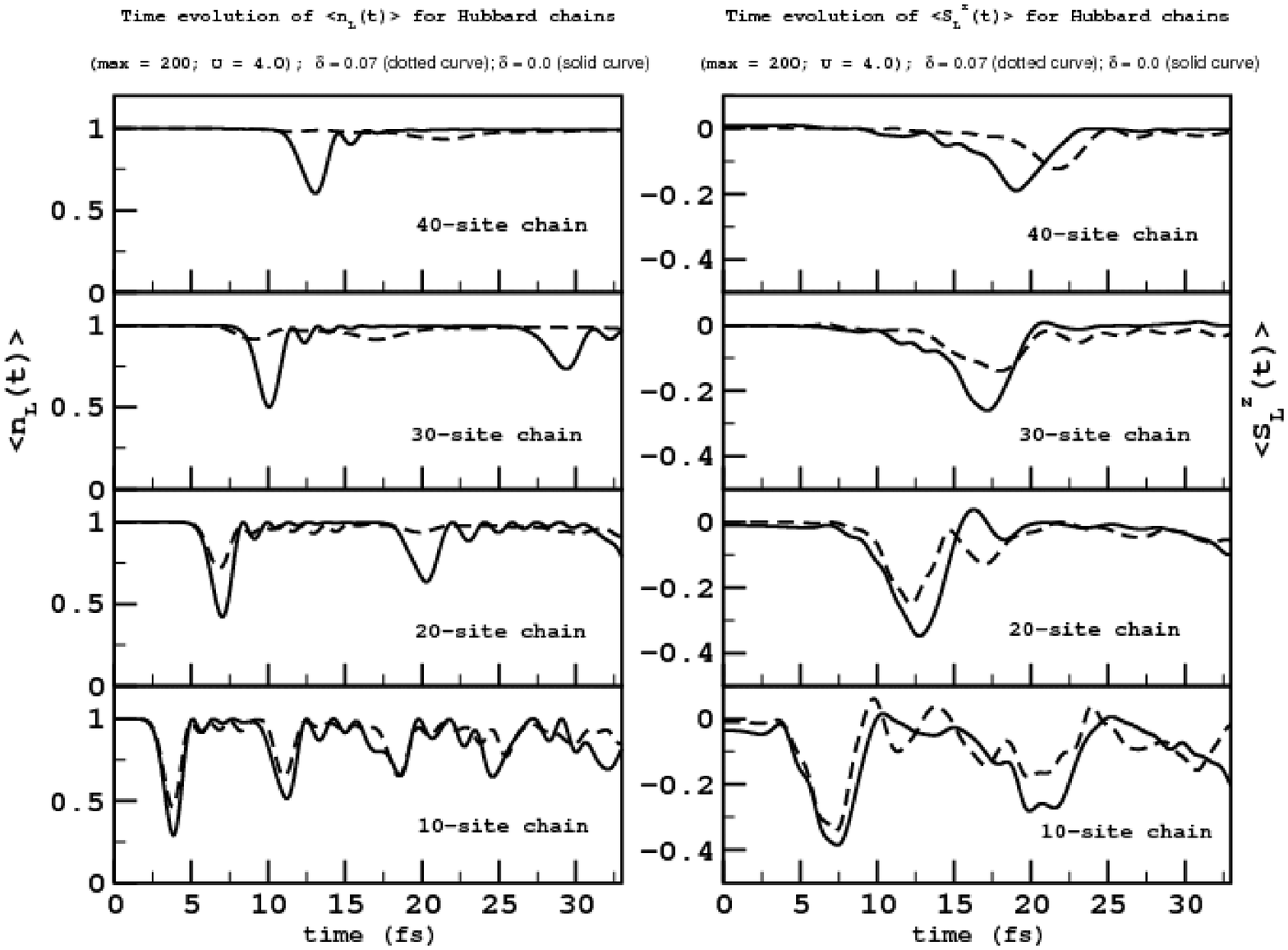}}
\caption{\small 
Time evolution of $\bra n_{L}(t) \ket$ ({\bf Left}) and $\bra S^{z}_{L}(t) \ket$ ({\bf Right}) for uniform
(solid curve) and dimerized (dashed curve) Hubbard chains of length 10, 20, 30 and 40 sites, with
$\frac{U}{\mid t \mid}$ = 4.0. Dimension of the DMEV basis used is 200.}
\label{figure-4}
\end{center}
\end{figure}

\vspace{0.2cm}
\begin{figure}[!htbp]
\begin{center}
{\includegraphics[scale=0.6]{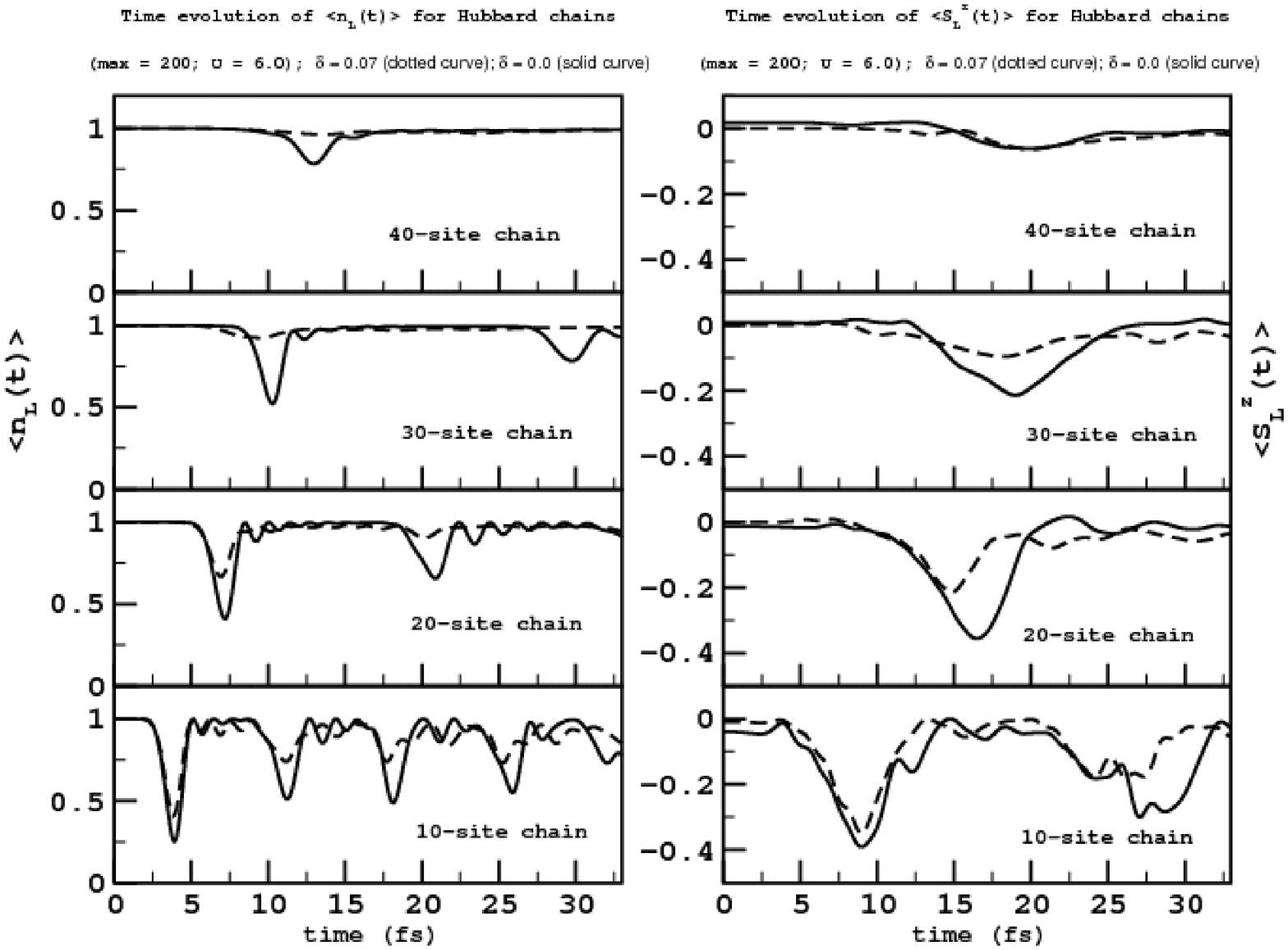}}
\caption{\small 
Time evolution of $\bra n_{L}(t) \ket$ ({\bf Left}) and $\bra S^{z}_{L}(t) \ket$ ({\bf Right}) for uniform
(solid curve) and dimerized (dashed curve) Hubbard chains of length 10, 20, 30 and 40 sites, with
$\frac{U}{\mid t \mid}$ = 6.0. Dimension of the DMEV basis used is 200.}
\label{figure-5} 
\end{center}
\end{figure}

\vspace{0.2cm}
\begin{figure}[!htbp]
\begin{center}
{\includegraphics[scale=0.6]{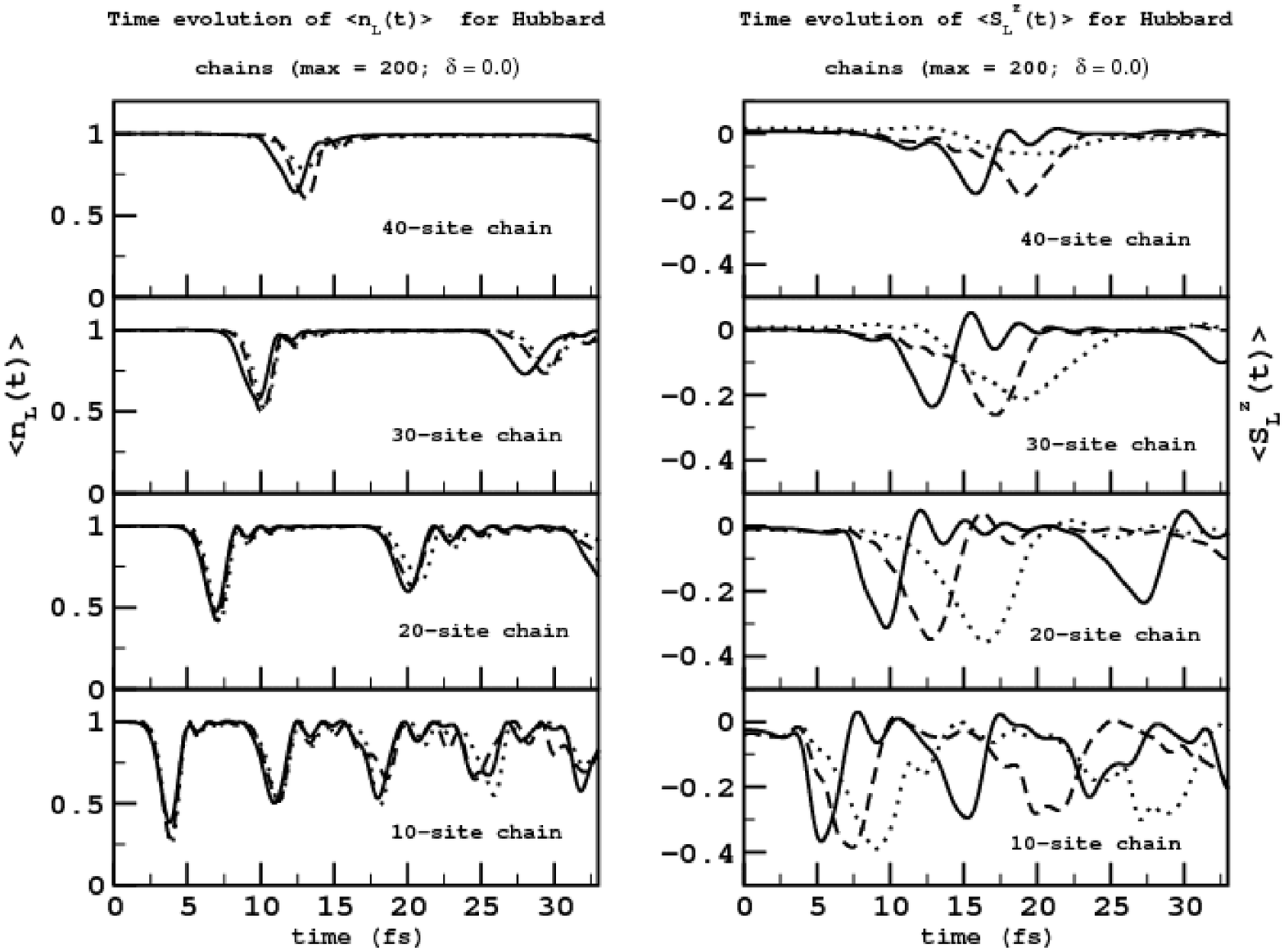}}
\label{figure-6}
\caption{\small 
Comparison of the time evolutions of $\bra n_{L}(t) \ket$ ({\bf Left}) and $\bra S^{z}_{L}(t) \ket$ 
({\bf Right}) for uniform Hubbard chains of length 10, 20, 30 and 40 sites, with
$\frac{U}{\mid t \mid}$ = 2.0 (solid curve), 4.0 (dashed curve) and 6.0 (dotted curve). 
Dimension of the DMEV basis used is 200.}
\end{center}
\end{figure}

\vspace{0.2cm}
\begin{figure}[!htbp]
\begin{center}
{\includegraphics[scale=0.6]{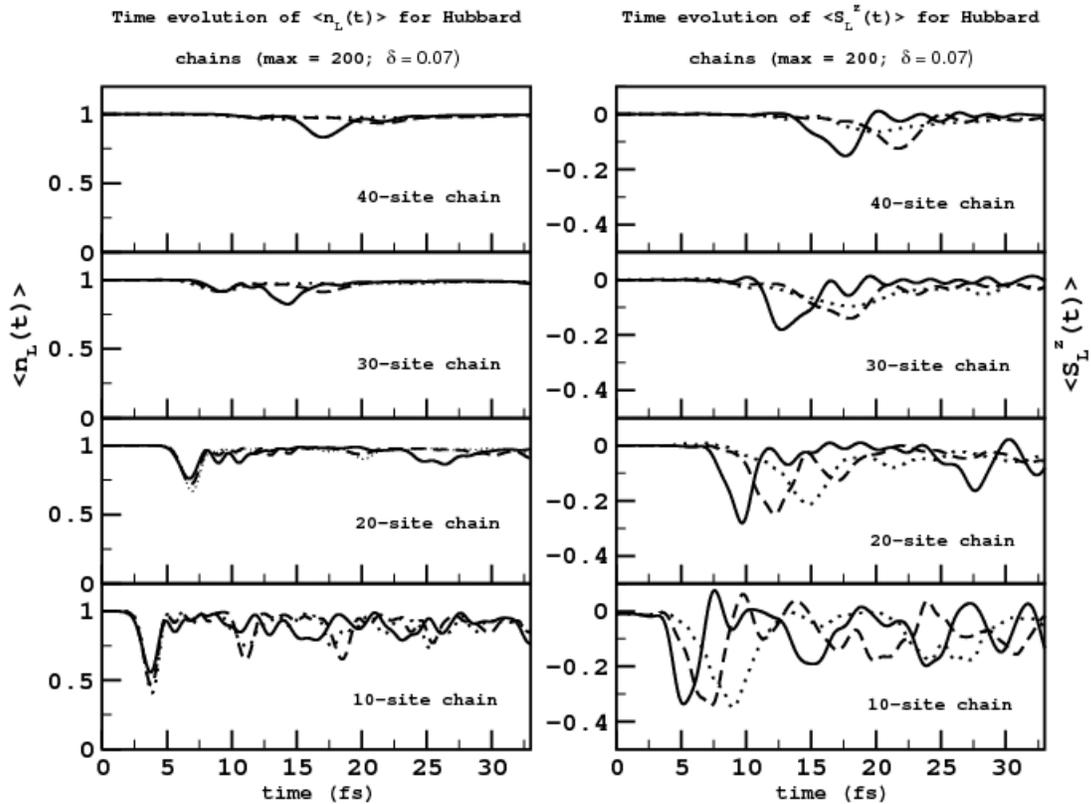}}
\caption{\small
Comparison of the time evolutions of $\bra n_{L}(t) \ket$ ({\bf Left}) and $\bra S^{z}_{L}(t) \ket$ 
({\bf Right}) for dimerized Hubbard chains ($\delta$ = 0.07) of length 10, 20, 30 and 40 sites, with
$\frac{U}{\mid t \mid}$ = 2.0 (solid curve), 4.0 (dashed curve) and 6.0 (dotted curve). 
Dimension of the DMEV basis used is 200.}
\label{figure-7}
\end{center}
\end{figure}

\newpage
\section{Summary and Outlook}
{\noindent 
To summarize, we find from our tdDMRG calculations that when a hole with a desired spin is injected 
at one end of the $\pi$-conjugated backbone of a polyene chain, it propagates from one end of the 
chain, to the other. The motion of this hole can be monitored by focusing our attention on the 
time evolution of the charge and spin densities at the two end-sites of the chain. In the absence 
of any external reservoirs (source-drain), the hole gets reflected back and forth across the length 
of the chain showing oscillatory motion. In the absence of electron-electron correlations, the 
charge and spin degrees of the hole do not get decoupled. The time taken by the hole to travel across 
the whole polyene backbone increases with approximately constant velocity. We are currently extending
these studies to PPP model and polymer topologies involving phenyl rings. It is seen that for  
dimerized chains, the velocity decreases further because of the fact that velocity of the hole 
is determined by the smaller of the two hopping matrix element, $t_{i,i+1} = t_{0}(1-(\delta)^{i})$, 
where $\delta$ is the dimerization parameter and $t_{0}$ is the mean hopping matrix element. 
For Hubbard chains, where spin-charge separation occurs, the hole "breaks-up" into two elementary 
excitations, one carrying only charge (holon), and the other, only spin (spinon), both of which moves 
with different velocities. It is found in accordance with the earlier literature, 
the holon moves faster than the spinon, and with increasing U, although the
velocity of the holon remains "almost" unaltered, that of the spinon significantly decreases.}
\vspace{0.5cm}

\noindent {\bf Acknowledgment}
This work was supported by grants from the department of science and technology 
(DST), India.

\end{document}